\documentstyle[epsfig]{mn}
\begin{document}

\title
[Statistical lensing by galactic discs]
{Statistical lensing by galactic discs}
\author
[A.\,W. Blain, Ole M\"oller \& Ariyeh H. Maller]
{
A.\,W. Blain,$^1$ Ole M\"oller$^1$ and Ariyeh H. Maller$^2$\\
$^1$ Cavendish Laboratory, Madingley Road, Cambridge, CB3 0HE.\\
$^2$ Physics Department, University of California, Santa Cruz, 
CA\,95064, USA.
}
\maketitle

\begin{abstract}
The high-magnification gravitational lensing cross-section of a galaxy is 
enhanced significantly if a baryonic disc is embedded in its dark matter 
halo. We investigate the effects of a population of such discs on the 
probability of detecting strongly lensed images of distant galaxies and 
quasars. The optical depth to lensing is always more than doubled. The effects 
are particularly significant at magnifications greater than about 50, for which the 
optical depth typically increases by a factor greater than 5. If either the 
fraction of discs or the typical disc-to-halo mass ratio increases with redshift, 
then the optical depth is expected to be increased even further. Obscuration 
by dust in the lensing disc is expected to counteract the associated enhanced 
magnification bias for surveys in the optical waveband, but not in the radio and 
millimetre/submillimetre wavebands. The presence of galactic discs should 
hence lead to a significant increase in the number of lensed galaxies and 
quasars detected in both radio- 
and millimetre/submillimetre-selected surveys. An increase by a factor of about 
3 is expected, enhancing the strong case for millimetre/submillimetre-wave lens 
surveys using the next generation of ground-based telescopes and the {\it 
FIRST} and {\it Planck Surveyor} space missions. 
\end{abstract}  

\begin{keywords}
galaxies: evolution -- galaxies: fundamental 
parameters -- cosmology: theory -- gravitational lensing -- 
radio continuum: galaxies 
\end{keywords}

\section{Introduction}

There has been much recent interest in the effects of galactic discs on 
the lensing cross-sections of galaxies (Maller, Flores \& Primack 1997; Wang \& 
Turner 1997; Bartelmann \& Loeb 1998; Keeton \& Kochanek 1998; 
Koopmans, De Bruyn \& Jackson 1998; M\"oller \& Blain 1998). A 
gravitationally dominant baryonic disc in the core of a dark matter halo, 
edge-on to the line of sight to a distant source, can significantly increase the 
mass surface density, and thus the lensing efficiency. 

The image geometry produced by an individual lens can be modified significantly 
by the presence of an edge-on or almost edge-on disc; however, the 
cross-section to multiple imaging is only increased by a factor of about 50\,per 
cent when averaged over all disc inclinations. In contrast, the cross-section 
to magnifications greater than 10 is expected to be increased by a factor of 
a few after the same averaging process (M\"oller \& Blain 1998). Hence, 
depending on the abundance and form of evolution of discs in dark matter 
haloes, the strong lensing optical depth at moderate and large redshifts could be 
increased very significantly. Bartelmann \& Loeb (1998) have also recently 
discussed this effect.
They note that it will probably not be possible to exploit any enhanced 
magnification in the optical waveband because the flux densities of the lensed 
images will be attenuated by dust associated with the disc. For the same reason, 
Maller et~al. (1997) did not consider the effect of disc lensing on the 
magnifications of multiple images because of possible differential extinction 
between the images. 

Dust extinction is important in the optical waveband; however, a dusty lensing 
disc is optically thin in the radio and millimetre/submillimetre wavebands. Indeed, 
because of $k$--correction effects (Blain \& Longair 1993), a lensing object 
is often expected to be fainter than the lensed images in the submillimetre 
waveband (Blain 1997a, 1998b). Hence, any enhanced magnification bias due 
to the effects of discs would be easy to detect in these wavebands. Existing 
radio lens surveys, such as the CLASS survey (Browne et al. 1998), appear 
to show an overabundance of lens candidates that can be identified as highly 
inclined discs. The potential importance of submillimetre-wave lens surveys 
has been discussed for lensing by both clusters (Blain 1997a, 1998a) and 
galaxies (Blain 1996, 1997b, 1998b,c). Submillimetre-wave lensing by clusters 
has now been demonstrated to be a very useful phenomenon (Smail, Ivison \& 
Blain 1997). The most striking feature of submillimetre-wave galaxy lens surveys 
is that the fraction of strongly lensed galaxies detected in a carefully selected 
sample can exceed 5 per cent (Blain 1996, 1998b).

In Section 2 we discuss the statistical features of lensing by galaxies 
containing a disc, using techniques and models of lensing by disc 
galaxies developed by M\"oller (1996), Maller et al. (1997) and M\"oller \& Blain 
(1998), and the statistical lensing formalism (Peacock 1982; Pei 1995) applied in 
the submillimetre waveband by Blain (1996). In Section\,3 we predict the effects 
of a potentially evolving population of discs (Lilly et~al. 1998; Mo, Mao \& White 
1998; Mao, Mo \& White 1998; Marzke et al. 1998) on the lensing cross-sections. 
In Section\,4 we estimate the surface density of strongly lensed 
galaxies and quasars in the light of recent submillimetre-wave observations 
by Smail et al. (1997) using the new SCUBA camera at the James Clerk Maxwell 
Telescope (Holland et al. 1999). See Smail et al. (1998, 1999, in preparation), 
Ivison et al. (1998), Blain, Ivison \& Smail (1998a), Blain et al. (1998b, 1999a,b)  
and Frayer et al. (1998) for a discussion of the latest results. In 
Section~5 the prospects for detecting such images in future surveys using 
forthcoming ground-based and space-borne instruments are discussed. 

The values of cosmological parameters have a significant effect on the 
properties of galaxy--galaxy submillimetre-wave lensing (Blain 1998a). 
Unless otherwise stated, we assume an Einstein--de Sitter world model 
with Hubble's constant $H_0 = 60$\,km\,s$^{-1}$\,Mpc$^{-1}$, in
accordance with the models presented by Maller et al. (1997) and M\"oller \& 
Blain (1998). Note that the abundance of submillimetre-wave lensed galaxies 
predicted in an Einstein--de Sitter model is less than that expected in a model 
with a smaller density parameter (Blain 1998a). 

\section{The strong lensing optical depth} 

\subsection{Basic formalism}

Millimetre/submillimetre-wave source counts of lensed 
images were predicted by Blain (1996), 
using formalism developed by Peacock (1982) and Pei (1995) to estimate 
the optical depth to strong lensing attributable to galaxies. 
In this formalism the probability that a source at redshift $z$ is lensed by a 
magnification between $\mu$ and $\mu + {\rm d}\mu$ is $F(\mu,z)$. This function 
is obtained by integrating the cross-sections of individual lenses to 
magnifications greater than $\mu$, $\sigma_\mu(M, z)$ in the plane of the
lens, both along the line of sight to $z$ and over the mass function of lenses 
$N$, and then differentiating with respect to $\mu$,
\begin{equation}
F = - { {\rm d} \over {{\rm d}\mu} } \int_0^z \int_0^\infty \sigma_\mu(M, z') 
(1+z')^2 {\rm d}N(M) \, { {{\rm d}r} \over {{\rm d}z'} }\,{\rm d}z'. 
\end{equation}
Here $r$ is the comoving radial distance element. The number and flux 
density conservation conditions,
\begin{equation}
\int_0^\infty F(\mu,z) \> {\rm d}\mu = 
\int_0^\infty \mu \, F(\mu,z) \> {\rm d}\mu = 1, 
\end{equation}
must be satisfied. If the intrinsic luminosity function of galaxies $\Phi(L,z)$ is 
known, then the source count of unlensed galaxies
\begin{equation}
N(S_\nu) = {1 \over {4\pi} } \int_0^{z_0} 
\int_{L_{\rm min}(S_\nu, z)}^\infty \Phi(L,z) \> {\rm d}L \> 
{ {D(0,z)^2} \over {(1+z)^2} } \,
{ { {\rm d}r } \over { {\rm d}z } } \, {\rm d}z.
\end{equation}
$L_{\rm min}$ is the luminosity of a source at a redshift $z$ that 
produces a flux density $S_\nu$; $z_0$ is the maximum redshift of 
the source population; and $D(z_{\rm a}, z_{\rm b})$ is the angular diameter 
distance between redshifts $z_{\rm a}$ and $z_{\rm b}$. The count of 
lensed galaxies can be determined using equation (3), if $\Phi$ is replaced 
by an effective luminosity function,  
\begin{equation}
\Phi'(L,z) = \int_0^{\mu_{\rm max}(z)} { {F(\mu,z)} \over \mu } \>
\Phi \left( {L \over \mu}, z \right) \> {\rm d}\mu,
\end{equation}
corrected to include the effects of lensing.
$\mu_{\rm max}$, the maximum magnification, is determined by the intrinsic size 
of the source and the geometry of the lens configuration. In general, larger 
values of $\mu_{\rm max}$ are expected for smaller sources. 
$\mu_{\rm max}(z=1)\simeq 25$ is expected for lensing by a singular 
isothermal sphere (SIS) lens if the intrinsic size of the source is 
1\,kpc (Peacock 1982). 

\subsection{Lensing by singular isothermal spheres (SISs)}     

The cross-section to high-magnification lensing by an SIS at redshift $z'$ for a 
source at redshift $z$, as measured in the plane of the lens,
\begin{equation}
\sigma_\mu^{\rm SIS} (\mu, M_{\rm e}) = 64\pi
\left[ { G \over {c^2} } { {M_{\rm e}} \over R } \right]^2
\left[ { { D(0,z')D(z',z) } \over {D(0, z)} } \right]^2 \mu^{-2}, 
\end{equation}
(Schneider, Ehlers \& Falco 1992). $M_{\rm e}$ is the mass of the lens enclosed 
within a radius $R$. The distance ratio inside the second square bracket is 
referred to as $D_{\rm R}(z', z)$ in subsequent equations. In the case of lensing 
by SISs, the magnification and redshift dependences of $F$ (equation 1) 
can be separated at large magnifications, yielding the strong lensing
probability $F_{\rm SIS} = a(z) \mu^{-3}$ that is applicable here. 
The form of $F_{\rm SIS}$ at all magnifications can be approximated by  
\begin{equation}
F_{\rm SIS} = H \delta(\mu - \mu_0) + \cases{a(z) \mu^{-3}, &
                        if $\mu_t \le \mu \le \mu_{\rm max}$;\cr
                                0, & otherwise. \cr
}
\end{equation} 
The first term describes the typically small magnification correction that 
must be applied to most sources in order to satisfy the number and flux 
density conservation conditions (equation 2).

The form of the function $a(z)$ depends on both the values of 
cosmological parameters and the form of evolution of lensing galaxies 
(Blain 1996, 1998a). If the distribution of lenses in mass $M$ is described by a 
Press--Schechter function (Press \& Schechter 1974),  
\begin{equation}
{\rm d}N(M,z) = { {\bar \rho_{\rm g}} \over {\sqrt\pi} } { \gamma \over {M^2} }
\left( {M \over {M^*} } \right)^{\gamma/2} 
{\rm exp}\left[ - \left({ M \over {M^*} }\right)^\gamma \, \right] \, {\rm d}M, 
\end{equation}
then equation (1) can be evaluated analytically. $M^*(z)$ includes all the 
details of the evolution of structure resulting from the growth of perturbations 
in an expanding Universe; $\gamma = 4/3$ if the primordial density 
fluctuations are scale-independent; $\bar\rho_{\rm g}$ is the mean smoothed 
density of bound objects in the Universe; and $M$ is the mass of a region of 
the Universe that has turned around from the Hubble flow and is in the 
process of collapse at a particular epoch. $M$ is related to the 
mass enclosed within a radius $R$, $M_{\rm e}$ (equation 5) by the 
constant $\epsilon$, that is,  
\begin{equation} 
M_{\rm e} = \epsilon M.
\end{equation}
$M^*_{\rm e}$ is the equivalently transformed version of $M^*$. This 
transformation allows the masses in equations (5) and (7) to be linked in 
the same expression. For example, if the population of lensing galaxies 
consists entirely of SISs, then by evaluating equation (1), 
\begin{eqnarray}
\lefteqn{\nonumber a(z) = 128 \sqrt{\pi} \bar\rho_{\rm g} \epsilon
\left[ { G \over { {c^2} R } } \right]^2
\Gamma \left[ { 1 \over \gamma} + { 1 \over 2 } \right] \times } \\
& & \>\>\>\>\>\>\>\>\>\>\>\>\>\>\>\>\>
\int_0^z  D_{\rm R}^2(z',z) \,(1+z')^2 M_{\rm e}^*(z') 
{ {{\rm d}r} \over {{\rm d}z'} }\,
{\rm d}z'.
\end{eqnarray}
If $F_{\rm SIS}$ is given by equation (6) then
\begin{equation}
\Phi'_{\rm SIS} = { H \over {\mu_0} } \, \Phi \left( { L \over {\mu_0} }\right) +
a(z) \int_{\mu_{\rm t}}^{\mu_{\rm max}(z)} { {I(\mu)} \over \mu } \> \Phi
\left( { L \over \mu } \right) {\rm d}\mu,
\end{equation}
in which $I(\mu) = \mu^{-3}$. This equation was used to derive the results 
for the probabilities of statistical lensing presented by Blain (1996, 1998a). 

\subsection{Lensing by discs}

M\"oller \& Blain (1998) recently presented the results of simulations of the 
high-magnification lensing cross-section for an exponential disc embedded 
within an SIS halo. When the disc was within about $30^\circ$ of being edge-on 
to the line of sight, this cross-section was found to be increased by a significant 
factor as compared with that for an SIS halo alone. An example of the relative 
magnification $\sigma_\mu / \sigma_\mu^{\rm SIS}$ is plotted in fig.\,3(a) of 
M\"oller \& Blain (1998), for a disc/halo combination well matched to that 
of the Milky Way. We have now extended these simulations to include a wide 
range of halo and disc masses, and so to investigate the statistical properties of 
lensing by such objects. The central surface mass density of the disc within the 
halo was kept constant (Mao et al. 1998), while the scalelength of the 
disc was varied in such a way that the ratio of the mass of dark matter in the SIS 
halo within a radius $R=10$\,kpc and the mass of the disc remained constant. 
The results are now parametrized to generate an analytical description of the 
effects of lensing by a population of discs. 

\subsubsection{Formalism}

The relative cross-section to lensing by a magnification greater than $\mu$ 
due to a combination of a disc and an SIS halo, as compared with an 
SIS halo alone, 
was approximated by a function of the form
\begin{equation}
{ {\sigma_\mu} \over {\sigma_\mu^{\rm SIS} } } \! = \! C(M_{\rm e})
\! \cases{ \!\displaystyle{
{{ (\beta-\alpha) {\mu_*}^{1+\alpha} \mu^2} \over 
{1 + \alpha + \beta + \alpha\beta} } - 
{ {\mu^{3+\alpha}} \over {1+\alpha}}}, & if $\mu \le \mu_*$;\cr
\!\displaystyle{ - { {{\mu_*}^{\alpha-\beta} \mu^{3+\beta}} \over {1+\beta}} }
, & otherwise, \cr
}
\end{equation}
where
\begin{equation}
C(M_{\rm e}) = K \left[ \, 1 + 
\left({ {M_{\rm e}} \over {M_0}}\right)^\eta \, \right].
\end{equation}
Any systematic evolution of the scalelength/mass of discs within dark matter 
haloes can be incorporated by modifying the value of $M_0$, defined as the 
mass of dark matter enclosed within 10\,kpc of the core of an SIS halo, as a 
function of redshift.

\begin{table}
\caption{The values of the best-fitting parameters derived from the extended 
results of M\"oller \& Blain (1998). The associated best-fitting curves are shown 
in Fig.\,1. The best-fitting values of the worst constrained parameters, 
$\beta$ and $\mu_*$, are related by the equation $\beta = 0.017\mu_* - 2.6$. 
}
{\vskip 0.75mm}
\hrule
{\vskip 1.2mm}
\begin{tabular}{ p{1.2cm} p{2.9cm} p{1.2cm} p{3.1cm} } 
Parameter & Value & Parameter & Value\\
\end{tabular}
{\vskip 1.2mm}
\hrule 
{\vskip 1.2mm}
\begin{tabular}{ p{1.2cm} p{2.9cm} p{1.2cm} p{3.1cm} } 
$\alpha$ & $-2.85 \pm 0.16$ & $\beta$ & $-2.12^{+0.25}_{-0.14}$\\
$\eta$ & $-0.87 \pm 0.06$ & $\mu_*$ & $26^{+14}_{-8}$\\
$M_0$ & $(2.6 \pm 0.4)\times 10^{11}$\,M$_\odot$ & $K$ & $1.7 \pm 0.5$\\
$M^*_{\rm e}$ & $2.9 \times 10^{11}$\,M$_\odot$ & & \\
\noalign{\vskip 1.5mm}
\end{tabular}
\hrule
\end{table}

Equation (11) has the useful property that when differentiated to obtain 
${\rm d}\sigma_\mu/{\rm d}\mu$ (required in equation 1), the result has a 
simple double-power-law form,
\begin{equation}
{ {\rm d \sigma_\mu} \over { {\rm d}\mu } } = 64\pi 
\left[ { G \over {c^2} } { M_{\rm e} \over R } \right]^2 
C(M_{\rm e}) \cases{\mu^\alpha, & if $\mu \le \mu_*$;\cr
{\mu_*}^{\alpha-\beta}\,\mu^\beta, & otherwise. \cr
}
\end{equation}

The parameters $\alpha$, $\beta$, $\mu_*$, $K$, $M_0$ and $\eta$ were 
determined by fitting the function in equation (11) to the results of the numerical 
simulations (M\"oller \& Blain 1998). The form of the best-fitting models, and 
probability contours determined by fitting to the worst-constrained pairs of 
parameters, $\beta$ and $\mu_*$, and $\eta$ and $M_0$, are all shown in Fig.\,1. 
The values of all the best-fitting parameters are listed in Table\,1.

\begin{figure*}
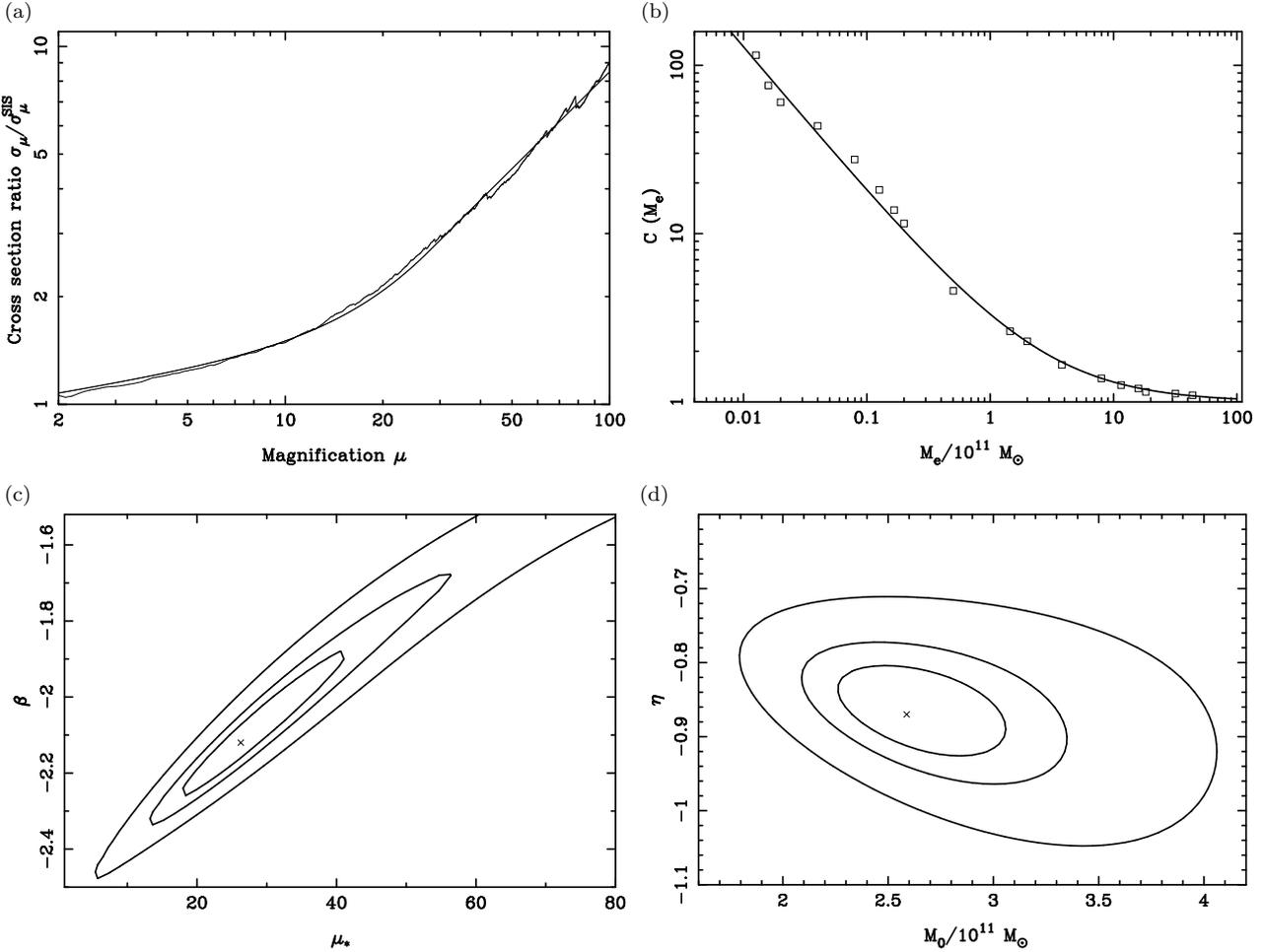

\begin{minipage}{170mm}
(a) \hskip 81mm (b)
\begin{center}
\vskip -4mm
\hskip -2mm \epsfig{file=fig1a.ps,width=5.9cm,angle=-90} \hskip 1mm
\epsfig{file=fig1b.ps,width=5.9cm,angle=-90} 
\end{center}
(c) \hskip 81mm (d)
\vskip -3mm
\begin{center}
\vskip -4mm
\hskip -2mm \epsfig{file=fig1c.ps,width=5.9cm,angle=-90} \hskip 1mm
\epsfig{file=fig1d.ps,width=5.9cm,angle=-90}
\vskip -2mm
\end{center}
\caption{The best-fitting models that describe (a) the magnification dependence 
of the function $\sigma_\mu / \sigma_\mu^{\rm SIS}$ (equation 11) and (b) the 
mass dependence of the function $C(M_{\rm e})$ (equation 12), shown by the 
smooth curves. The results of the simulations are shown by the jagged line in (a) 
and the empty squares in (b) respectively. 
The quality of the constraints imposed on 
the worst constrained pairs of parameters in each case are shown in (c) and (d) 
respectively. The three contour levels correspond to likelihoods 1$\sigma$, 
2$\sigma$ and 3$\sigma$ away from the maximum value, which is marked by
a cross. The values of the best-fitting parameters are listed in Table\,1.}
\end{minipage}
\end{figure*} 

The resulting model for the cross-section to strong lensing by discs within 
SIS haloes can be used to estimate the optical depth to strong lensing 
$F_{\rm disc}$. $F_{\rm disc}$ is obtained as a function of redshift by evaluating 
equation (1), incorporating the fraction of haloes that contain a baryonic disc 
$f_{\rm d}(z)$: 
\begin{eqnarray}
\lefteqn{\nonumber F_{\rm disc} = - { {\rm d} \over {{\rm d}\mu} } \int_0^z 
\int_0^\infty (1+z')^2 \, { {{\rm d}r} \over {{\rm d}z'} } \>\>\> \times} \\
& & \>\>\>\>\>\>\>\>\>\>\>
\left[ \, (1-f_{\rm d}) \sigma_\mu^{\rm SIS}(M) + f_{\rm d} \sigma_\mu(M) \, \right]
\, {\rm d}N(M) \,{\rm d}z'.
\end{eqnarray}
For magnifications $\mu > \mu_{\rm t}$ this can be written as, 
\begin{equation}
F_{\rm disc} = \left[ a_1(z) - a_2(z) \right] I_{\rm n}(\mu) + 
\left[a_2(z) + a_3(z) \right] I_{\rm d}(\mu),     
\end{equation}
the sum of two magnification-dependent terms, $I_{\rm n}$ and $I_{\rm d}$, 
which are modulated by redshift-dependent terms that are combinations of 
three functions, $a_1$, $a_2$ and $a_3$. As in equation (6), a low-magnification 
correction term containing $H$ and $\mu_0$ must be included in order to 
approximate the complete probability function. 

$I_{\rm n}$ and $I_{\rm d}$ are associated with lensing by pure SIS haloes and 
by discs within SIS haloes respectively: 
\begin{equation}
I_{\rm n}(\mu) = \mu^{-3}, 
\end{equation}
and
\begin{equation}
I_{\rm d}(\mu) = {K \over 2} 
\cases{ \mu^\alpha, & if $\mu \le \mu_*$;\cr
           \mu_*^{\alpha-\beta} \mu^\beta, & otherwise. \cr
}
\end{equation}
The $a$-functions (equation 15) are 
\begin{equation} 
a_1(z) = \Gamma\left[ {1 \over \gamma} + {1 \over 2}\right] 
\int_0^z J(z') M^*_{\rm e}(z') \,{\rm d}z', 
\end{equation}
\begin{equation} 
a_2(z) =  \Gamma\left[ {1 \over \gamma} + {1 \over 2}\right] 
\int_0^z J(z') f_{\rm d}(z') M^*_{\rm e}(z') \,{\rm d}z', 
\end{equation}
and 
\begin{equation} 
a_3(z) = \Gamma\left[{ {\eta+1} \over \gamma } + {1 \over 2 }\right]
\int_0^z \!\!\!\! J(z') 
{ { {M^*_{\rm e}(z')}^{(1+\eta)} } \over { M_0(z')^{\eta} } } 
f_{\rm d}(z')\,{\rm d}z',
\end{equation}
in all three of which
\begin{equation}
J(z') = 128 \sqrt{\pi} \bar\rho \epsilon \left[ { G \over { {c^2} R } } \right]^2 
D_{\rm R}^2(z', z) \> (1+z')^2 { {{\rm d}r} \over { {\rm d}z' } }.  
\end{equation}
Note that if discs do not evolve, then the redshift dependences of $a_1$, $a_2$ 
and $a_3$ are identical. 

The relative normalization of $a_3$ as compared with $a_1$ and $a_2$ depends 
on the ratio of the mass $M_0$, derived from the results of the numerical 
simulations described above, and the typical mass in the Press--Schechter 
function $M^*_{\rm e}$ (equations 7 and 8). The value of $M^*_{\rm e}$ can be 
determined by using the Tully--Fisher relation (Hudson et al. 1998), an 
empirical power-law relation between the $B$-band luminosity and circular 
velocity of galaxies. A circular velocity $v_{\rm c} \simeq 360$\,km\,s$^{-1}$ is 
indicated by the Tully-Fisher relation for an $L^*$ galaxy with a $B$-band 
magnitude $M_B = -21$ at the present epoch. If $L^*$ is assumed to be 
equivalent to $M^*$, then $M^*_{\rm e}$ can be determined by comparison with 
the Milky Way, for which $M_{\rm e}(R<10\,{\rm kpc})=10^{11}$\,M$_\odot$ and 
$v_{\rm c}\simeq 210$\,km\,s$^{-1}$. 
$M_{\rm e}^*\simeq 2.9\times10^{11}$\,M$_\odot$ is expected in this case. 

By combining equations (15) to (21), we can model the optical depth to strong 
lensing by a population of SIS dark matter haloes that contain an evolving 
population of baryonic discs.
A new effective luminosity function $\Phi'_{\rm disc}$ can be produced by 
including the general form of $F_{\rm disc}$ in equation (4). In order 
to evaluate the count of lensed sources using this function, the forms of 
$a_1(z)$, $a_2(z)$, $a_3(z)$, $f_{\rm d}(z)$ and $M_0(z)$ are required. The
redshift dependences of the $a$-functions in both hierarchical and
non-hierarchical models, and three different world models, are illustrated 
in Fig.\,2. The form of evolution of discs, discussed by Lilly et al. (1998) and 
Mao et al. (1998), is considered in Section~3.1. 

We parametrize the fraction of haloes that contain discs $f_{\rm d} = f_0 (1+z)^p$ 
and the scalelength $r_{\rm s} \propto (1+z)^{q/2}$. The comoving abundance 
of discs and the mass of the disc within a typical halo are thus expected to 
evolve as $M^*(z)^{-1} f_{\rm d}(z)$ and $M^*(z) (1+z)^{q}$ respectively, if the 
central surface density of the disc is assumed to be independent of redshift. In an 
Einstein--de Sitter model, $M^* \propto (1+z)^{-1}$ would be expected 
because of the hierarchical evolution of bound objects. 

\begin{figure*}
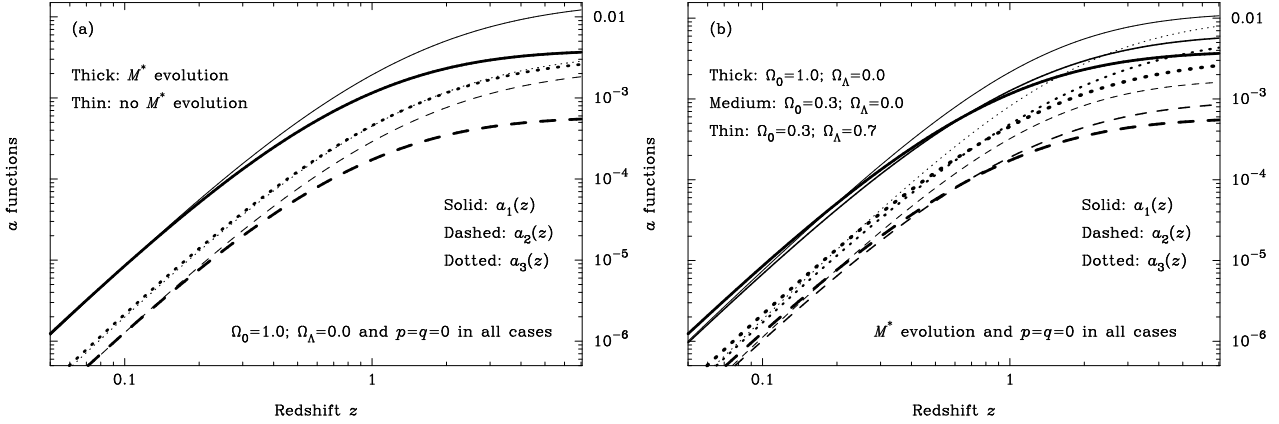

\begin{minipage}{170mm}
\begin{center}
\epsfig{file=fig2a.ps,width=5.5cm,angle=-90} \hskip 1mm
\epsfig{file=fig2b.ps,width=5.5cm,angle=-90}
\end{center}
\caption{A comparison of the three redshift-dependent $a$-functions that  
contribute to $F_{\rm disc}$ (equation 15). In (a) the effects of 
including evolution of $M^*$ in the Press--Schechter function 
are shown; $a_1$ and $a_2$ are affected in the same way, but the effect 
on $a_3$ is much less significant. In (b) $M^*$ is assumed to evolve, and 
the effects of three different world models are compared: see also Blain (1998a). 
The functions are normalized to match the low-redshift value of $a(z)$ used in 
Blain (1996), with a value of $f_0=0.15$: see Section\,3.1.}
\end{minipage} 
\end{figure*}

\subsubsection{Normalization of the terms in $F_{\rm disc}$}

The relative size of the two terms in equation (15) depends on the parameters 
$f_0$, $p$, $q$ and $\bar\rho\epsilon$. To clarify the dependence of
the results on the form of evolution, we introduce the parameters $\tau$ and 
$R_{\tau}$, the total probability of lensing by a magnification $\mu > \mu_{\rm t}$, 
and the ratio of the probabilities of lensing by a disc as compared with lensing by 
a halo alone respectively: 
\begin{equation}
\tau = [a_1-a_2] \int_{\mu_{\rm t}}^\infty \!\!\!\! I_{\rm n}(\mu)\,{\rm d}\mu + 
[a_2+a_3] \int_{\mu_{\rm t}}^\infty \!\!\!\! I_{\rm d}(\mu)\,{\rm d}\mu, 
\end{equation}
and
\begin{equation}
R_\tau = { {a_2+a_3} \over {a_1-a_2} } \> 
{ {\int_{\mu_{\rm t}}^\infty I_{\rm d}(\mu)\,{\rm d}\mu} \over 
{\int_{\mu_{\rm t}}^\infty I_{\rm n}(\mu)\,{\rm d}\mu } }.  
\end{equation}
At a low redshift $z_1$, for which all three $a$-functions are proportional to 
$z^3$, and for values of $\mu_{\rm t} \simeq 2$ then
\begin{eqnarray}
\tau &\simeq& 
16 \sqrt{\pi} \bar\rho \epsilon \left[ { G \over { {c^2} R } } \right]^2
(1 + f_0T) \> \Gamma\left[ {1 \over \gamma} + {1 \over 2} \right] \times \nonumber \\
& & \>\>\>\>\>\>\>\>\>
\int_0^{z_1} 
D_{\rm R}^2(z', z) \> (1+z')^2 M^*_{\rm e}(z') 
{ {{\rm d}r} \over { {\rm d}z' } }\,{\rm d}z',
\end{eqnarray}
in which
\begin{equation}
T = 
{ { \Gamma\left( {{\eta+1}\over\gamma} + {1 \over 2} \right) } \over 
{ \Gamma \left( {1 \over \gamma} + {1 \over 2} \right) } }
\left( { {M^*_{\rm e}} \over {M_0} } \right)^\eta, 
\end{equation}
and
\begin{equation}
R_\tau \simeq { {f_0} \over {1-f_0} } \, ( 1 + T).
\end{equation}
Based on the values of parameters listed in Table\,1, $T$ is expected to 
take the value $1.6$. 
If $\tau \simeq 10^{-5}$ at $z=0.1$ (Blain 1996), then the density parameter 
associated with $\bar\rho_{\rm g}$ is 
$\Omega_{\rm g} \epsilon \simeq 8 \times 10^{-4}$. Hence, since 
$\Omega_{\rm g} \sim 10^{-2}$, $\epsilon \sim 8 \times 10^{-2}$. This would 
indicate that 8\,per cent of the mass of the dark matter in the SIS halo of a 
$M^*$ galaxy at the present epoch was enclosed within 10\,kpc of the core, 
equivalent to a truncation radius of 125\,kpc for the SIS.

If $f_0=0.6$ (see Section 3.1), then $R_\tau \simeq 3.6$, and so about 80 per cent 
of lensing events should be caused by disc galaxies, neglecting the effects of 
both magnification bias and evolution of the population of lenses. 

By combining equations (24) to (26), the low-redshift normalization of the 
$a$-functions can be determined by comparison with the normalization of $a(z)$ 
in a disc-free model (equation 6): see Blain (1996) for a summary of the 
estimated form of $a(z)$. The normalization of $a_1$ is given by
\begin{equation}
a_1(z_1) = {1 \over 2} {\tau_{\rm n}(z_1) \mu_{\rm t}^{-2} }\, (1+f_0T), 
\end{equation}
where $\tau_{\rm n}$ is the probability of lensing in a model without discs, 
to which the predictions in Blain (1996, 1998a, 1998c) were normalized. 
The other two $a$-functions are normalized to 
\begin{equation}
a_2(z_1) = f_0 a_1(z_1),
\end{equation}
and
\begin{equation}
a_3(z_1) = f_0 T a_1(z_1).  
\end{equation} 
 
\begin{figure*}
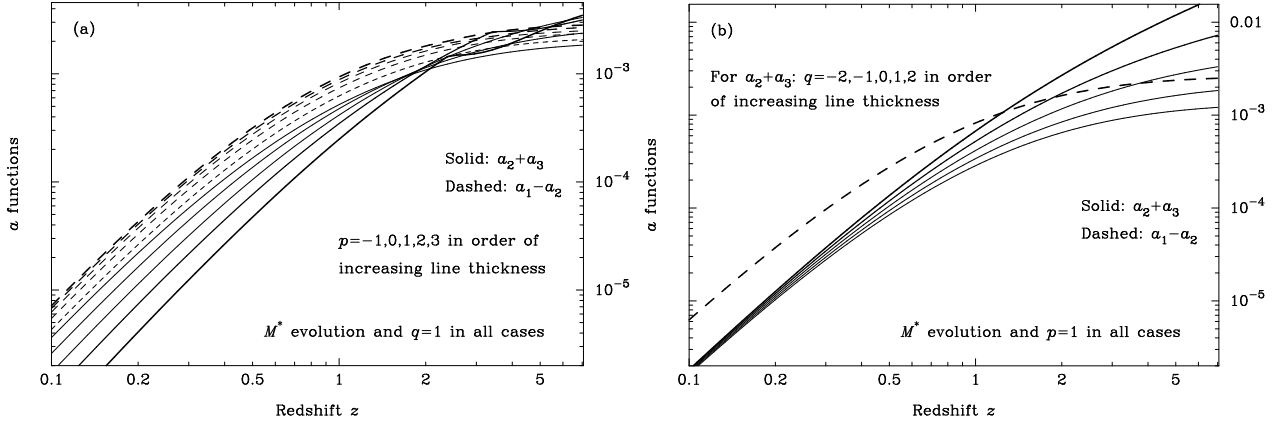

\begin{minipage}{170mm}
\begin{center}
\epsfig{file=fig3a.ps,width=5.5cm,angle=-90} \hskip 1mm
\epsfig{file=fig3b.ps,width=5.5cm,angle=-90}
\end{center}
\caption{A comparison of the effects of disc evolution on the two 
redshift-dependent functions that contribute to $F_{\rm disc}$ (equation 15), 
$a_1-a_2$ and $a_2+a_3$. In (a) five different forms of the abundance of discs 
$f_{\rm d}(z) = f_0 (1+z)^p$ are compared, with $p=-1$, 0, 1, 2 and 3; the
disc-to-halo mass ratio does not evolve. In each case, the value of $f_0$ is 
chosen to satisfy the constraint that the fraction of discs found in the CFRS 
$R_{\rm d} \simeq 0.18$ (equation 30). In (b) the effects of five different forms of 
the evolution of $M_0 \propto (1+z)^q$ are included, with $q=-2$, $-1$, 0, 1 and 
2; a value of $p=1$ is assumed in all cases, corresponding to a value of 
$f_0=0.089$ if $R_{\rm d}=0.18$. The discontinuities in the gradients of the solid 
curves at high redshifts in (a) are caused by $f_{\rm d}$ reaching its upper limit. 
Note that the strongly evolving models have values of $f_0$ that are much
smaller the observed zero-redshift value (Marzke et al. 1998).  
}
\end{minipage} 
\end{figure*}

\section{Lensing cross-section predictions}

\subsection{The population and evolution of discs}

Lilly et al. (1998) discussed the morphological properties of galaxies 
detected in the Canada--France Redshift Survey (CFRS). To avoid surface 
brightness selection effects, they classified galaxies based on their appearance 
within 4\,$h^{-1}$\,kpc of the core. They found no evidence for evolution in 
the abundance of spiral galaxies out to redshifts greater than 0.5. About 18\,per 
cent of the 
galaxies with identifiable morphologies in the small-radius sample contained a 
disc. Based on a $B$-band selected local galaxy survey, unaffected by 
surface brightness selection effects, Marzke et al. (1998) find that about 64\,per 
cent of local galaxies have a disc morphology. Brinchmann (private
communication) find that 
about 50\,per cent of galaxies selected in the $I$ band are classified 
as containing a disc, either by eye or by machine. The fraction does not appear 
to depend  strongly on either the redshift or the intrinsic luminosity of the 
galaxies. The zero-redshift abundance of discs in haloes, $f_0$, thus appears to 
be about 0.6. 

Based on the results of theoretical and numerical work on the evolution of 
discs (Mo et al. 1998), Mao et al. (1998) argued that the results
presented by Lilly et al.\ (1998) are consistent with a comoving abundance of 
discs that varies as $(1+z)^3$, and a typical scalelength 
$r_{\rm s} \propto (1+z)^{-1}$. If $M^*$ does not evolve, then values of $p=0$ 
and $q=0$ correspond to no evolution of either the abundance or the typical 
scalelength of discs (Lilly et al. 
1998); if $M^*$ does evolve, then values of $p=-1$ and $q=1$ would be required. 
If the abundance and scalelength of discs evolves as indicated by Mao et al. 
(1998), then values of $p=2$ and $q=-1$ are required if $M^*$ evolves; values 
of $p=3$ and $q=-2$ are required if $M^*$ does not evolve.

Several physical processes can be invoked to explain the evolution of discs. 
If discs grow by the accretion of baryons from their haloes, then the disc-to-halo 
mass ratio would be expected to be smaller at large redshifts. Disk--disc
or disc--halo mergers would generally disrupt discs, reducing
the abundance of discs at low redshifts. Low-mass discs could also be 
disrupted by supernova-driven winds associated with episodes of powerful 
star formation. A determination of the evolution of the properties of discs 
would thus have important consequences for understanding the formation and 
evolution of galaxies. 

The dominant contribution to the lensing cross-section is provided by 
galaxies at redshifts less than unity. Hence, the results of Lilly et al. (1998), 
which sample this redshift range, should provide useful information about the 
normalization constant $f_0$. In the absence of selection effects, the fraction of 
discs, 
\begin{equation}
R_{\rm d} \simeq
{\displaystyle
{\int_0^1 f_0(1+z')^p \, { {D(0,z')^2} \over { (1+z')^2 } } 
M^*(z')\, {{{\rm d}r}\over{{\rm d}z'}} \, {\rm d}z'} \over 
{\displaystyle
\int_0^1 \left[ 1 - f_0 (1+z')^p \right] { {D(0,z')^2} \over { (1+z')^2 } } 
M^*(z')\, {{{\rm d}r}\over{{\rm d}z'}} \,{\rm d}z'}
},
\end{equation}
subject to the constraint that $f_0 (1+z')^p \le 1$. If $p=0$, then 
$R_{\rm d} (\simeq 0.18) = f_0 / (1 - f_0)$, and so $f_0 \simeq 0.15$.
If the disc fraction increases with increasing redshift, that is $p>0$, then the 
value of $f_0$ will be reduced. The form of evolution of $M^*$ has very little 
effect on the results. For values of $p=-1$, 0, 1, 2 and 3, $f_0$ takes the 
values of 0.25, 0.15, 0.089, 0.051 and 0.025 respectively. 

\begin{figure*}
\begin{minipage}{170mm}
\begin{center}
\hskip -7.83cm
\epsfig{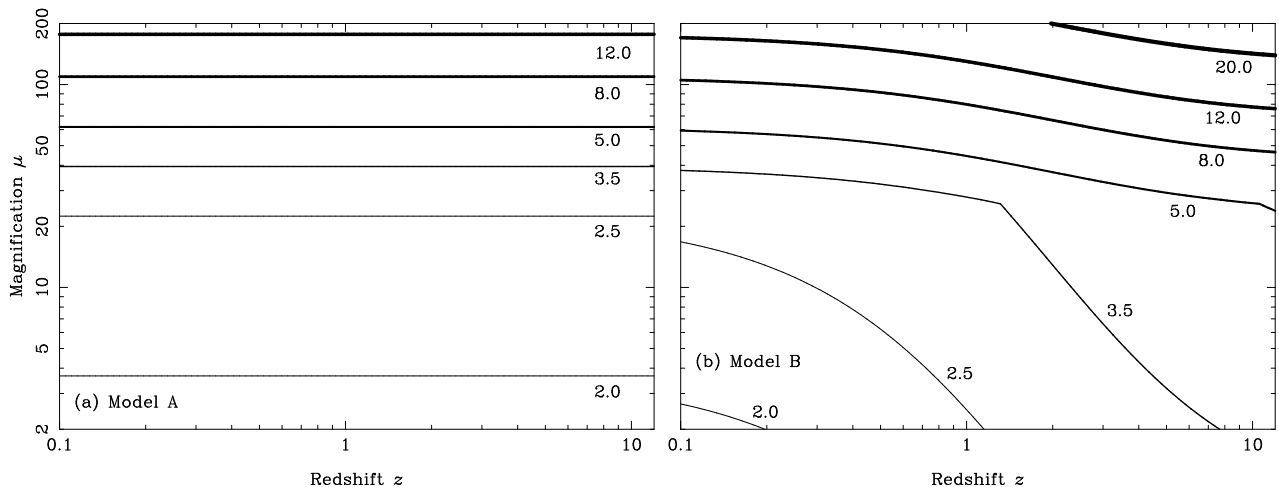}
\end{center}
\caption{The ratio of the differential optical depths to high-magnification 
lensing produced by a population of discs within SIS haloes and a population 
of pure SIS haloes, $F_{\rm disc} / F_{\rm SIS}$ as a function of redshift and 
magnification. Models A and B, which include different values of the parameters 
$p$ and $q$, are defined in Table\,2. The lensing parameters listed in Table\,1 
are assumed. The results are not affected significantly by changes to the 
world model.  
}
\end{minipage}
\end{figure*}

In the absence of universal agreement on the interpretation of the CFRS data, 
we present forms of the functions $a_1$, $a_2$ and $a_3$ derived for a range 
of different values of $p$ and $q$ to illustrate the possible effects of different 
forms of disc evolution: see Fig.\,3. The most significant effects are
produced by large values of $q$. These values correspond to models in which 
the disc-to-halo mass ratio evolves strongly. 
Large lens surveys carried out in 
the radio and millimetre/submillimetre wavebands should eventually provide a 
large sample of mass-selected lensed objects. These samples could be used 
to resolve any remaining uncertainty in the interpretation of existing surveys. 

\subsection{Forms of $F_{\rm disc}$ derived for evolving discs}

\begin{table}
\caption{The values of the parameters in the disc lensing models used to derive
the results in Figs 4 and 5. Both models are constrained by requiring that they 
provide a reasonable, if schematic, account of the abundance of discs determined 
in the CFRS fields by Lilly et al. (1998), normalized to the disc fraction at 
zero redshift found by Marzke (1998). In both models, discs have the same 
properties at all redshifts less than unity (Lilly et al. 1998). 
See Section\,2.3.1 for a definition of the parameters $f_0$, $p$ and $q$.   
}
{\vskip 0.75mm}
\hrule
{\vskip 1.2mm}
\begin{tabular}{ p{1.1cm} p{2.5cm} p{1.2cm} p{1.2cm} p{1.2cm} } 
Model  & $M^*$ evolution? & $f_0$ & $p$ & $q$ \\
\end{tabular}
{\vskip 1.2mm}
\hrule 
{\vskip 1.2mm}
\begin{tabular}{ p{1.1cm} p{2.5cm} p{1.2cm} p{1.2cm} p{1.2cm} } 
A & No & 0.60 & 0 & 0 \\
B & Yes & 0.60 & -1 & 1 \\
\noalign{\vskip 1.5mm}
\end{tabular}
\hrule
\end{table}
 
The optical depths to strong lensing $F_{\rm disc}$ and $F_{\rm SIS}$ can now 
be calculated and compared.  In Fig.\,4 $F_{\rm disc}/F_{\rm SIS}$ is shown for 
both models of disc evolution listed in Table\,2 as a function of both 
magnification and redshift. The optical depth to strong lensing is increased by 
incorporating the effect of lensing by discs by a factor between 2 and 3 
at redshifts of about unity and magnifications of several tens. The effect of 
including discs is more significant at greater magnifications. The consequences 
for the counts of lensed galaxies and quasars in the submillimetre waveband 
are discussed below. 

\section{Lensed galaxies in the submillimetre waveband} 

\subsection{Introduction} 

The selection functions in submillimetre-wave surveys are very broad and
extend out to high redshifts. This is because the flux density--redshift relation 
for distant dusty galaxies is expected to be flat over the range of redshifts from 
about 0.5 to 10, owing to large negative $k$--corrections. These arise when dust 
emission spectra, peaking in the rest-frame far-infrared waveband, are 
redshifted into the submillimetre waveband (Blain \& Longair 1993). A
submillimetre-wave survey thus selects high-redshift galaxies preferentially, 
and because a flat flux density--redshift relation naturally corresponds to
a steep source count, magnification biases in such a survey are expected to be 
large (Blain 1996). At that time, however, source counts of submillimetre-wave 
sources had not been determined. Now Smail et al.\ (1997) have detected a 
population of distant dusty galaxies, including a dust-shrouded starburst/AGN 
at a redshift $z=2.803$ (Ivison et al.\ 1998; Frayer et al.\ 1998). A fuller description 
of the survey (Smail et al. 1998, 1999, in preparation; Blain et al. 1999b) and 
its consequences (Blain 1998c; Blain et al. 1998a,b, 1999a) can be found 
elsewhere. The results of other submillimetre-wave surveys are now becoming 
available (Barger et al. 1998; Eales et al. 1998; Holland et al. 1998; 
Hughes et al. 1998). 

The surface density of galaxies with flux densities greater than 4\,mJy at a 
wavelength of 850\,$\mu$m is $(2.5 \pm 1.4) \times 10^3$\,deg$^{-2}$ 
(Smail et al.\ 1997). Counts of lensed galaxies that are consistent with these 
observations, which do not include the effects of lensing by discs, were 
presented by Blain (1998a,b,c). The local population of {\it IRAS} galaxies 
(Saunders et al. 1990) appears to undergo pure luminosity evolution of the form 
$(1+z)^\gamma$ where $\gamma \simeq 4$ out to $z \simeq 2.5$ (Blain et al. 
1999a). This model is consistent with counts at other wavelengths in the 
millimetre/submillimetre waveband (Wilner \& Wright 1997; Kawara et al. 1998; 
Lagache et al. 1998), and with the intensity of diffuse background radiation in 
the submillimetre/far-infrared waveband (Puget et al. 1996; Guiderdoni et al. 
1997; Schlegel, Finkbeiner \& Davis 1998; Hauser et al. 1998; Fixsen et al. 1998).

\subsection{Calculating counts of lensed galaxies} 

We determine counts of lensed galaxies by using equation (3) with $\Phi'$
calculated using  $F_{\rm disc}$ (equation 15), normalised as described above, 
in both models of disc evolution described in Table\,2. The population of 
distant dusty galaxies assumed is described by the `Gaussian' model of Blain et 
al. (1999a), which accounts for all the available submillimetre-wave and 
far-infrared data currently available.

In order to illustrate the expected form of the counts of both lensed and
unlensed galaxies at wavelengths of 1300, 850, 450 and 60\,$\mu$m, 
the counts of galaxies expected in the absence of discs are compared in 
Fig.\,5(a). Lensed counts are calculated, assuming that $M^*$ does and 
does not evolve. If $M^*$ does not evolve, then the predicted count of 
lensed galaxies is greater. As a result, all previous predictions have been made 
assuming that $M^*$ evolves to produce conservative estimates of the counts of
lenses. To demonstrate the effects of including lensing by discs, the
relative size of the counts of lensed galaxies with discs as compared with 
the counts shown in Fig.\,5(a) are shown in Fig.\,5(b). The additional bias arising 
from disc lensing is typically a factor of 2 and 3 in models A and B respectively. 

The only data currently available on the population of lensed dusty galaxies
and quasars is provided by the identification of two luminous distant 
60-$\mu$m {\it IRAS} sources as lenses (Close et al. 1995; Irwin et al. 1998; 
Lewis et al. 1998), and 
the identification of one known lensed quasar with a 60-$\mu$m {\it IRAS} 
source (Kneib et al. 1998). This indicates a 60-$\mu$m count of lenses of at 
least $7 \times 10^{-5}$\,deg$^{-2}$ at a flux density of 0.2\,Jy, fully 
consistent with the value of about $7 \times 10^{-4}$\,deg$^{-2}$ shown in 
Fig.\,5(a). 

The maximum magnification assumed in these calculations for a source at 
$z=1$ is $\mu_{\rm max}(1) = 40$, as assumed by Blain (1996). A larger value of 
$\mu_{\rm max}$ will lead to a more significant effect 
of lensing by discs, as discs are progressively more efficient at producing large 
magnifications as compared with SIS haloes: see Figs\,1(a) and 4. This effect is 
emphasized in Fig.\,6, in which the additional magnification bias arising from 
lensing by discs is compared for different values of $\mu_{\rm max}(1)$: 10, 20, 
40 and 100. Observations made using millimetre-wave interferometers (Solomon 
et al. 1997; Mirabel et al. 1998; Sakamoto et al. 1999) 
indicate that the most luminous far-infrared emitting 
regions of distant galaxies are rather small, with scales of less than several 
hundred parsecs. As $\mu_{\rm max}(1)\simeq25$ is expected for a 1-kpc source 
(Peacock 1982), $\mu_{\rm max}(1)=40$ seems to be very reasonable. Larger 
magnifications may also be possible, at least for any very compact subsample 
of distant submillimetre-luminous galaxies and quasars, such as the `extreme 
starbursts' imaged by Downes \& Solomon (1998). In this case, the  
numbers of lenses with flux densities of order 100\,mJy, and thus the 
magnification biases due to discs (Fig.\,5b), will be increased.  

\begin{figure}
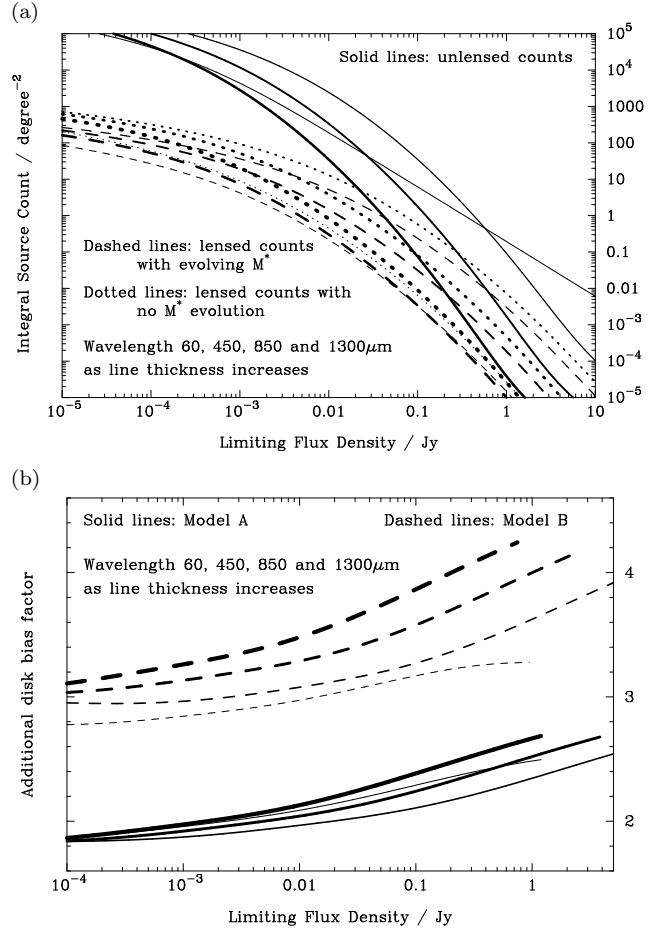

(a)
\begin{center}
\vskip -5mm
\epsfig{file=fig5a.ps, width=5.7cm,angle=-90}
\end{center}
(b)
\begin{center}
\vskip -4mm
\epsfig{file=fig5b.ps, width=5.7cm,angle=-90}
\end{center}
\caption{(a) Counts of lensed galaxies expected in the millimetre/submillimetre 
and far-infrared wavebands (Blain et al. 1999a). No effects of lensing by discs 
are included. Lensed counts with and without an evolving population of
haloes, as parametrized by $M^*$, are shown. (b) The additional magnification bias
factor that should multiply the counts in (a) when lensing by discs is 
considered in both models of disc evolution (Table\,2 and Fig.\,4). Some
curves stop before the right-hand edge of the figures because the predicted
counts fall below $2.4\times10^{-5}$\,deg$^{-2}$ ($1/4\pi$\,sr$^{-1}$).
}
\end{figure}

\begin{figure}
\begin{center}
\epsfig{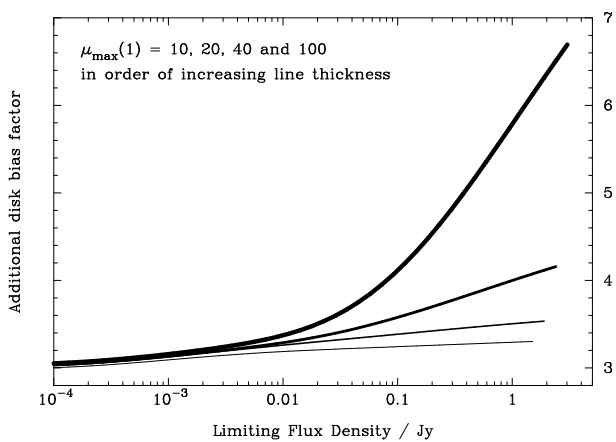}
\end{center}
\caption{Differences in the additional magnification bias due to lensing by discs, 
expected at a wavelength of 850\,$\mu$m in disc evolution Model\,B: see 
Table\,2. A larger value of $\mu_{\rm max}$ corresponds to a more significant 
effect, emphasizing the results shown in Fig.\,4.
}
\end{figure}

The fraction of lensed galaxies detected in a carefully designed flux-limited 
submillimetre-wave survey is expected to be several per cent (Blain 1996,
1998b), and it is reasonable to expect lensing by discs to increase this 
fraction by a factor of between 2 and 3. 

\section{The prospects for future observations}

\subsection{Potential instruments}

Many forthcoming instruments will be useful for carrying out 
submillimetre-wave lens surveys. Foremost amongst them are large 
ground-based millimetre/submillimetre interferometer arrays (MIAs), such as the 
MMA (Brown 1996): see also Downes (1996) and Ishiguro et al. (1992). 
These instruments, with subarcsecond angular resolution, will be required in 
order to confirm submillimetre-selected lens candidates (Blain 1998b). 
Large single-antenna telescopes such as the US--Mexican 50-m Large 
Millimeter Telescope (LTM/GTM; Schloerb 1997) and a 10-m telescope at the 
South Pole (Stark et al. 1998), are also being designed and built. These 
telescopes will provide large and uniform focal planes, able to exploit large 
format bolometer array receivers (Glenn et al. 1998), and would be very useful 
for wide-field millimetre/submillimetre-wave surveys (Blain 1998c). The 
{\it Planck Surveyor} (Bersanelli et al. 1996) and {\it Far-Infrared and 
Submillimetre Telescope} ({\it FIRST}; Pilbratt 1997) space missions 
also have great potential. 

\begin{figure*} 
\begin{minipage}{170mm}
(a) \hskip 56mm (b) \hskip 53mm (c)
\begin{center}
\vskip -3mm
\hskip -11.5cm
\epsfig{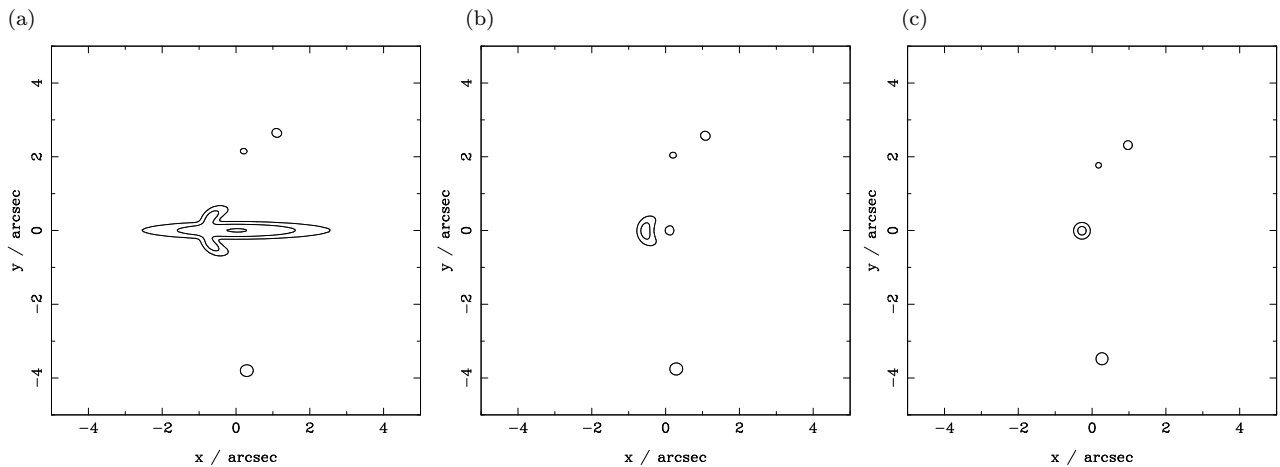}
\end{center}
\caption{
Simulated MMA images obtained at a wavelength of 850\,$\mu$m for a 
representative 
region of sky that is chosen to contain a strongly lensed source. About 2\,per 
cent of fields centred on a foreground galaxy would be expected to contain 
such a background source. The lensed images produced by a Milky Way 
galaxy at $z=0.3$, inclined 5$^\circ$ from the edge-on aspect, and the emission 
from the disc are shown in (a). In (b) the corresponding images produced by an
SIS lens with the same mass as the disc, also at $z=0.3$ are compared. The 
background population of sources are shown in (c). The contour levels 
correspond to the 1$\sigma$ noise levels expected in 1-min, 1-h and 
24-h integrations: 220, 28 and 5.8\,$\mu$Jy\,beam$^{-1}$ respectively. 
}   
\end{minipage}
\end{figure*} 

\subsection{The detection rate of lenses}

The relative speed of detecting lenses using the instruments discussed above
is considered in more detail elsewhere (Blain 1998c). There is freedom to 
choose a survey strategy for all but {\it Planck Surveyor}, and, if optimized, 
detection rates of order 0.03-0.3\,h$^{-1}$ are expected. If the effect of 
lensing by discs is included, then these detection rates would be expected to 
increase by a factor of about 2 or 3. If the optical depth to submillimetre-wave 
lensing by discs is as large as we expect, then follow-up observations of 
detected lenses should reveal that a large fraction are lensed by edge-on discs. 
Submillimetre-wave point sources will be extracted from the all-sky {\it Planck 
Surveyor} survey (Blain 1998b), down to the confusion limit (Blain et al. 1998a,b). 
More than a thousand lensed galaxies and quasars were predicted, without 
considering the effects of lensing discs. This number should also be increased 
by a factor of 2 or 3 when lensing by discs is taken into account.

\subsection{Imaging lens candidates using the MMA}

Multiple lensed images formed by galaxies are separated on arcsecond scales. 
A large interferometer array, such as the MMA, would thus be the ideal telescope 
with which to investigate lensing by distant galaxies in detail. In Fig.\,7 we show 
simulated images of a 10-arcsec wide field, contained within the primary 
beam of the MMA, at an angular resolution of 0.1\,arcsec. The contour levels 
correspond to three times the 850-$\mu$m 1$\sigma$ sensitivity limits in MMA 
integrations lasting 1\,min, 1\,h and 1\,d: 220, 28 and 5.8\,$\mu$Jy\,beam$^{-1}$ 
respectively. 
Note that the primary beam area of the MMA at 850\,$\mu$m
is about $3\times10^{-5}$\,deg$^{-2}$, and so the brightest source in this area 
would be expected to have a flux density of 40\,$\mu$Jy 
(Fig.\,5a). In Fig.\,7(c) a typical region of the background sky is shown. The
source distribution in this case is chosen so that one source will be magnified
significantly by a disc galaxy at the centre of the figure. All sources are 
Gaussian and 0.2\,arcsec across at full width at half-maximum. 
An alignment between 
source and lens this close or closer is expected in about 2\,per cent of cases.  
In Fig.\,7(a) the effects of lensing these sources by an edge-on disc galaxy, 
similar to the Milky Way, at a redshift of 0.3 is shown. The submillimetre-wave 
emission from the lensing galaxy is calculated by transforming the observation 
of a nearby edge-on disc galaxy (Israel, van der Werf \& Tilanus 1998) to $z=0.3$, 
giving a total flux density of about 1\,mJy and a scalelength of 1\,arcsec. In 
Fig.\,7(b) the much less significant effects of lensing by an SIS dark matter halo 
with the same mass at the same redshift are compared. 

In addition to following up the sources detected in surveys using
single-antenna telescopes (Blain 1998c), a deep targeted MIA survey of known
distant edge-on discs, provided for example by the Sloan Digital Sky Survey
catalogue (Margon 1998), would be a very powerful way of detecting lensed 
images of very faint distant submillimetre-wave galaxies. Integration times of 
about 24\,h per source would be required however. 

\section{Conclusions} 

\begin{enumerate}
\item We have estimated the enhanced magnification bias expected for statistical 
lensing due to a population of baryonic discs in dark matter haloes. The bias is
expected to be enhanced by a factor of 2 or 3 
as compared with a population of 
SIS lenses. This bias could be much larger if the sources are smaller than
several hundred parsecs in size, and can thus be magnified by a factor in excess
of 40. 
\item In the submillimetre waveband, lensed images will not suffer from the 
effects of dust absorption in the lensing galaxy. Hence images lensed by 
edge-on discs should be selected in an unbiased manner in submillimetre-wave 
surveys made using the MMA, LMT/GTM, a 10-m telescope at the South Pole, 
and the space-borne instruments {\it Planck Surveyor} and {\it FIRST}. 
\item The subarcsecond angular resolution of the MMA will be required in 
order to study lensed images in detail. A targeted search of known edge-on
discs could provide a fruitful technique for detecting and observing any very faint,
distant dusty galaxies.
\end{enumerate} 

\section*{Acknowledgements}

We thank Jarle Brinchmann, Malcolm Longair, Priya Natarajan, Joel Primack 
for helpful comments on the manuscript, the referee for his/her prompt 
reading of the manuscript and the Rotary Club of Erice 
for their hospitality. Observations made by Ian Smail, Rob Ivison, Jean-Paul 
Kneib and AWB have assisted greatly in refining the conclusions of this paper.

\end{document}